# Depth thermography: non-invasive 3D temperature profiling using infrared thermal emission

**Yuzhe Xiao[1], Chenghao Wan[1,2], Alireza Shahsafi[1], Jad Salman[1], and Mikhail A. Kats[1,2,3*]**

[1]*Department of Electrical and Computer Engineering, University of Wisconsin-Madison, Madison, Wisconsin 53706, USA*

[2]*Department of Materials Science and Engineering, University of Wisconsin-Madison, Madison, Wisconsin 53706, USA*

[3]*Department of Physics, University of Wisconsin-Madison, Madison, Wisconsin, USA 53706*

*corresponding author, email address: mkats@wisc.edu*

## Abstract:

We introduce a technique based on infrared thermal emission, termed *depth thermography*, that can remotely measure the temperature distribution beneath the surface of certain objects. Depth thermography utilizes the thermal-emission spectrum in the semitransparent spectral region of the target object to extract its temperature as a function of depth, in contrast with conventional thermography, which uses the spectrally integrated thermally emitted power to measure the surface temperature. Coupled with two-dimensional imaging, e.g., using an infrared hyperspectral camera or scanning a single-pixel spectrometer, this technique can yield volumetric temperature distributions. We carried out a proof-of-concept experiment on an asymmetrically heated fused-silica window, extracting the temperature distribution throughout the sample. Depth thermography may enable noncontact volumetric temperature measurements of microscopic objects such as multilayer electronic devices or macroscopic volumes of liquids and gasses as well as simultaneous all-optical measurements of optical and thermal properties.

## Introduction

Thermal emission is a ubiquitous and fundamental phenomenon by which energy in the form of electromagnetic waves is emitted due to temperature-dependent thermal fluctuations. The radiated fields from a surface depend on both the temperature and the optical properties of that surface, typically encoded in a parameter called the emissivity (or emittivity) [1]. Therefore, the thermal emission from an object can be used to measure its temperature if the emissivity is known or can be estimated. This is the basis for infrared thermometry, in which temperature is measured by integrating the thermal-emission power within the bandwidth of an infrared detector (typically around 8-14 μm) [2]. A two-dimensional (2D) version of this technology is infrared thermography, in which an infrared camera captures spatial thermal-emission information, producing a 2D image where intensity is mapped to the temperature of the object on its surface [3], [4]. Infrared thermography has been widely employed for remote sensing [5]–[7], medical imaging [8], and condition monitoring for machinery (because temperature is a common indicator of structural health) [9].

There are scenarios in which it is useful to measure the temperature profile throughout the volume of an object, rather than just on the surface. For example, thermal detection of localized hot spots caused by stress within the volumes of glasses and plastics during formation or extrusion can help improve production processes [10]. Knowing the temperature profiles of multilayer semiconductor devices such as light-emitting diodes [11] and quantum-cascade lasers [12], [13] can guide engineering efforts to improve their performance. Similarly, three-dimensional (3D) temperature imaging of chip-stack architectures can help better understand and manage thermal dissipation [14]. There are also cases where monitoring volumetric temperature of liquids and gases is important, such as in molten-salt nuclear reactors [15] and combustion engines [16]. One way to obtain such 3D temperature information is to put physical temperature probes into different locations throughout the volume, but this invasive approach may require probes that can withstand high-temperature or corrosive environments and may also perturb the system in question.

Non-invasive 3D temperature profiling has been demonstrated at microwave frequencies, via a technique called multi-frequency microwave radiometry [17]. This technique has been used to measure internal temperatures of humans and other biological objects [18]–[20], and also to profile the temperature of the atmosphere as a function of altitude [21]. However, this method has intrinsic shortcomings. For example, typical objects at room temperature have thermal-emission distributions that peak at mid-infrared wavelengths, with much less signal in the microwave region [22]. Furthermore, this method has a lateral resolution on the order of a few centimeters because of the diffraction limit, so resolving temperature profiles of micro-scale objects is not feasible.

In this paper, we propose a new method, termed *depth thermography*, to extract 3D temperature profiles using thermal emission at infrared wavelengths. By focusing on the infrared semitransparent regions of thermally emitting objects where emission from different depths contributes to the total emitted spectrum, we can extract temperature as a function of depth from a measured spectrum. We expect a better signal-to-noise ratio compared to measurements at microwave frequencies, allowing for higher temperature accuracy. Moreover, by shrinking the wavelength from centimeters to microns, we expect a ~10,000-fold improvement in lateral resolution. As a proof of concept, we experimentally demonstrated this technique using an asymmetrically heated fused-silica window. First, we directly measured the emission spectrum from the heated window using a Fourier-transform spectrometer (FTS). Then we developed a theoretical model of the experiment and a retrieval algorithm to extract the underlying temperature profile. We propose depth thermography as a noninvasive, high-resolution technique that can profile the temperature distribution of volumes in three dimensions.

## Concept

The emitted thermal power depends on the temperature and optical properties of the emitting object. In the limit of very high optical loss, the penetration depth of light (*i.e.,* the depth at which the intensity of light drops to $1/e$ of its initial value) becomes much smaller than the physical size of the object, and the object is completely opaque. Therefore, in this limit, an emission measurement only probes the surface, as expected in conventional thermography. In contrast, when there is no optical loss, the object is completely transparent and does not emit at all, in accordance with Kirchhoff's law of thermal emission [23]. In the intermediate case where the material has some optical loss but not too much (*i.e.,* it is semitransparent; simple examples include silica in the spectral region of 3–8 μm [24] and calcium fluoride in 7–12 μm [25]), the penetration depth of light may be comparable to the physical size of the object, and thermally emitted light generated at various depths can escape to free space.

Depth thermography utilizes the thermal-emission spectrum in the semitransparent spectral region of the target object to extract the temperature as a function of depth. Figure 1 illustrates the basic idea. Thermal emission originating from the top surface ($z = 0$) escapes to free space without any attenuation. Thermal emission originating inside the material will be attenuated before it escapes to free space, with the degree of attenuation depending on the depth and the wavelength-dependent optical loss.

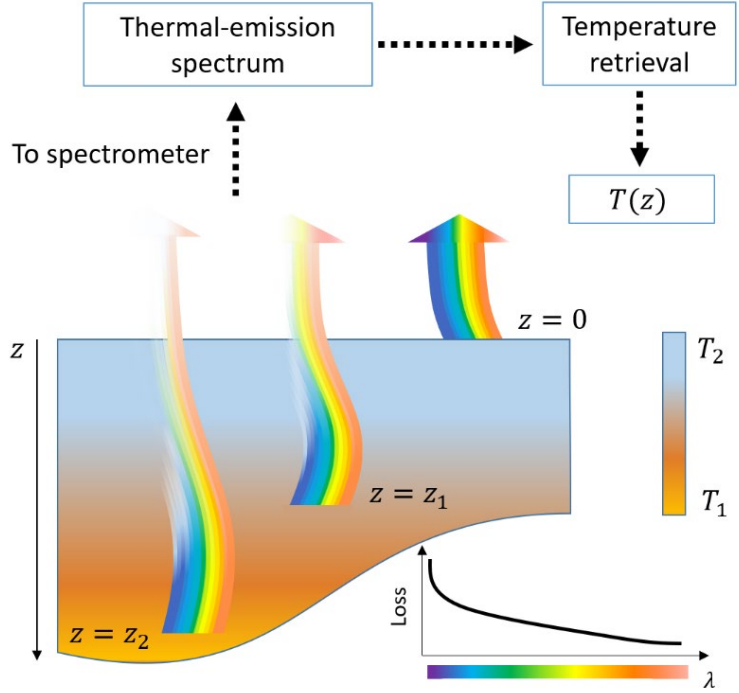

**Figure 1**. Basic concept of depth thermography. Thermal emission from an object with a non-uniform temperature distribution is considered. Light that originates inside the material is attenuated by the time it reaches the surface, with a spectral dependence that depends on the absorption spectrum of the material. In this figure, we consider the case of decreasing absorption vs. wavelength; therefore, shorter-wavelength components of the thermally emitted light are absorbed much more than the longer-wavelength components. A retrieval algorithm can be used to extract the temperature distribution beneath the surface using the measured thermal-emission spectrum.

The total thermal-emission spectrum comprises contributions from different depths, which are determined both by the local temperature and the optical properties throughout of the object. The local temperature and optical properties determine the intensity of the local thermal-emission source according to the fluctuation–dissipation relation [26], while the surrounding geometry and optical properties determine the transmission of thermal emission from different depths into free space. The transmission efficiency of light emerging from different depths into free space can be viewed as a series of eigenstates in the frequency domain, while the local temperature can be viewed as the corresponding eigenvalues. Finding the local temperature is therefore similar to the projection of the total thermal-emission spectrum onto these eigenstates. Below we describe our temperature-retrieval algorithm to extract the temperature profile $T(z)$ at different depths $z$ from the thermal-emission spectrum $I(\lambda)$ measured at the surface. Using this algorithm, a 3D temperature distribution can be obtained by laterally scanning the surface with a spectrometer or using a hyperspectral camera.

## Experimental results

Our experiment is diagrammed in Fig. 2(a). A 1-mm-thick fused-silica window was placed on the top of a heater surface, heated to 100, 200, or 300 °C. The relatively low thermal conductivity of fused silica ($k \sim 1.4$ W/m·K [27]) created a significant temperature gradient between the bottom surface that was in contact with the heater and the top surface that was exposed to air. The spectra of thermal emission emerging from the top of the fused-silica window were measured by sending the light into an FTS (Bruker VERTEX 70) [28], focusing on the 3–8 $\mu m$ spectral region where fused silica is semitransparent [24]. The sample was rotated by 10° with respect to the beam path to avoid multiple reflections between the sample and the interferometer [28]. For comparison, we also measured corresponding thermal-emission spectra from a laboratory blackbody reference: a ~500-$\mu$m-tall vertically-aligned carbon nanotube (CNT) forest [29], with an emissivity of ~0.98 across the mid infrared, which we have previously verified [28].

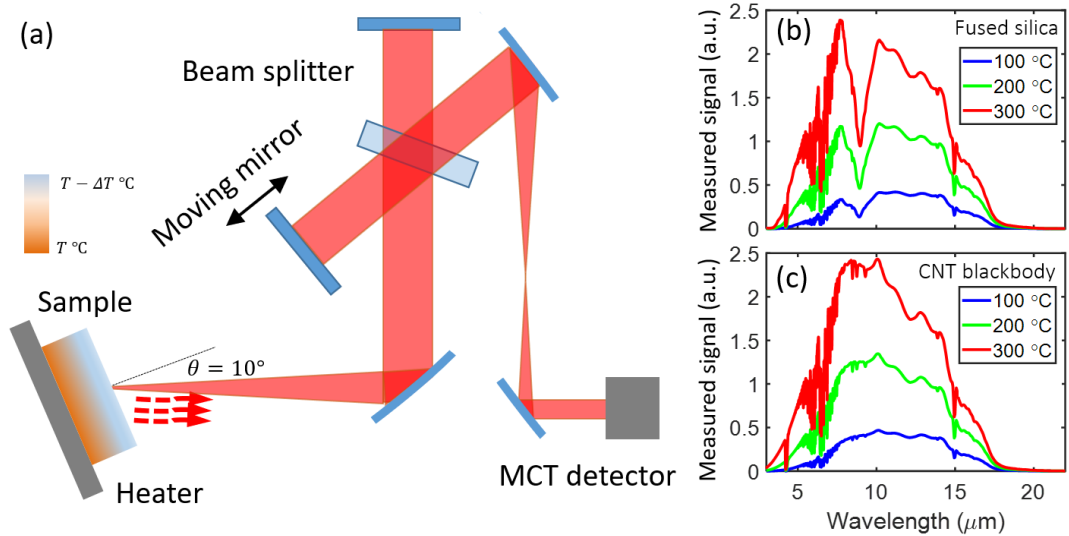

Figure 2. (a) Our thermal-emission measurement setup. The sample was placed on the top of a heated stage. Due to the finite thermal conductivity of the sample, there was a natural temperature gradient from the bottom to the top surface. The corresponding thermal emission was measured using a Fourier-transform spectrometer (FTS). (b) and (c) show the raw measured (but AC coupled and then Fourier-transformed) signal from a 1-mm thick fused-silica window and a laboratory CNT blackbody reference when the heater temperature was set to 100 (blue), 200 (green), and 300 °C (red).

Figure 2(b, c) shows the experimentally measured thermal-emission signal for the fused-silica window and the CNT blackbody reference at 100, 200, and 300 °C, respectively. The measured data from the samples were inevitably mixed with the background emission from components in the spectrometer beam path and the surrounding environment, making the data somewhat difficult to interpret [28]. This is further complicated by the possibility of a sample-dependent background signal. For example, there could be background emission from the surrounding

environment reflected by the sample into the FTS beam path. In general, the measured signal from the emitter $x$, $S_x(\lambda)$, can be expressed as:

$$S_x(\lambda) = m(\lambda)[I_x(\lambda) + B_x(\lambda)], \tag{1}$$

where $I_x(\lambda)$ is the true emission signal from sample $x$. In the case of thermal equilibrium (*i.e.* the emitter has a single uniform temperature $T$), $I_x(\lambda, T) = \epsilon_x(\lambda, T)I_{BB}(\lambda, T)$, where $\epsilon_x(\lambda, T)$ is the emissivity for emitter $x$ at temperature $T$ (in Kevin) and $I_{BB}(\lambda, T)$ is the blackbody-radiation distribution given by Planck's law [22]:

$$I_{BB}(\lambda, T) = \frac{2hc^2}{\lambda^5} \frac{1}{e^{\frac{hc}{\lambda k_B T}} - 1} \tag{2}$$

$B_x(\lambda)$ is the background, which can be wavelength- and sample-dependent, and $m(\lambda)$ is the wavelength-dependent system-response function that captures the collecting efficiency of the beam path and the detector response. To separate the true sample emission $I_x(\lambda)$ from the background $B_x(\lambda)$, we performed very careful calibration of our FTS (Sec. S1, Supplementary Information).

Figure 3(a, b) shows the calibrated thermal-emission spectra $I(\lambda)$ from the fused-silica window and the CNT blackbody at 100, 200 and 300 °C, respectively. The theoretical predictions based on the sample emissivity and Planck's law at the corresponding *stage* temperatures are plotted in the same figure. The emissivity of the fused-silica window was measured indirectly via Kirchhoff's law for opaque, flat samples [23] (i.e., $\epsilon(\lambda, T) = 1 - R(\lambda, T)$, where the measured reflectance $R(\lambda, T)$ is shown in Fig. S2, Supplementary Information). At $T = 100$ °C, the experimental spectra overlap reasonably well with the predictions due to Planck's law. However, the experimental spectra at higher temperatures (e.g., 200 and 300 °C) are smaller than the corresponding predictions for both the fused-silica window and the CNT blackbody. This observation is expected because samples were heated from the bottom and the thermal-emission spectra were measured from the top. The top surface temperature was lower than that of the heater stage and this difference was more pronounced for higher temperatures.

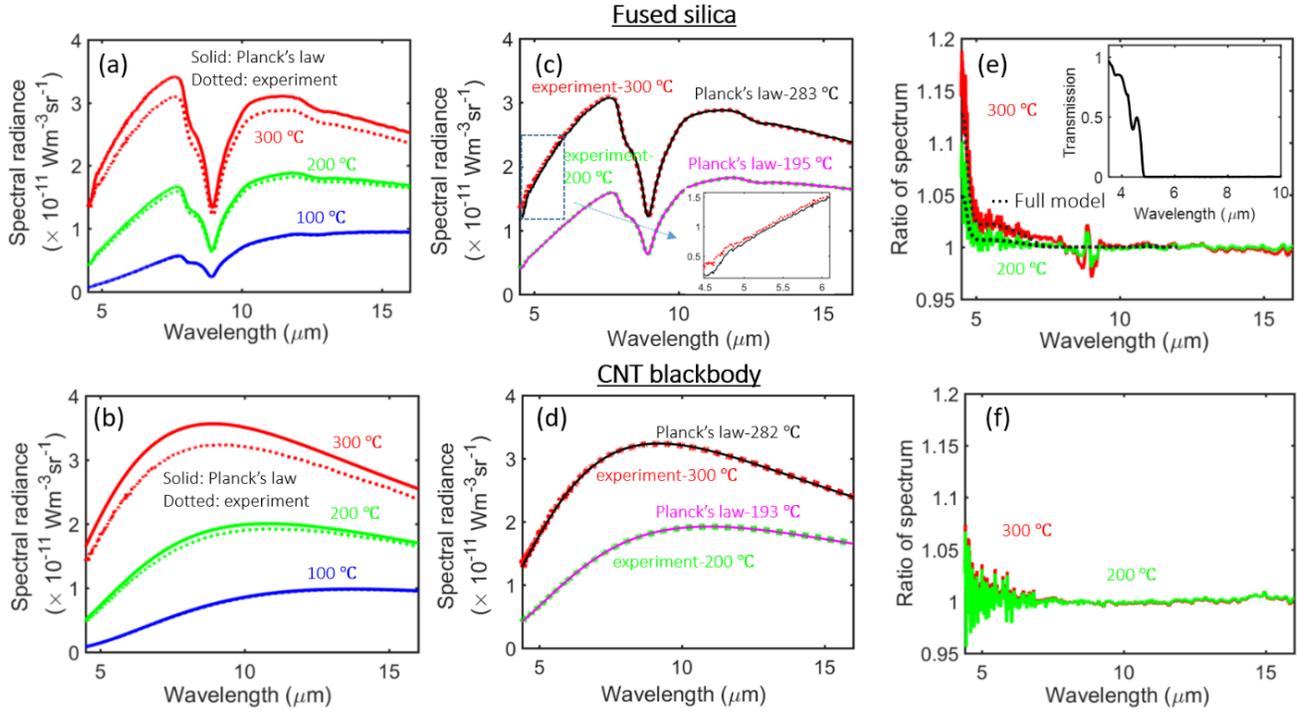

**Figure 3**. (a) Calibrated experimental thermal-emission spectra of the fused-silica window at 100 (dotted blue), 200 (dotted green), and 300 °C (dotted red). The theoretical predictions with the corresponding temperatures are plotted using solid lines in the same color. (b) The same as (a), but for the CNT blackbody. (d) The measured spectra of the CNT laboratory blackbody at 200 and 300 °C (dotted) can be fitted very well by Planck's law predictions at 193 and 282 °C (solid). (c) For λ > 8 μm, the measured spectra of the fused-silica window at 200 and 300 °C (dotted) can also be fitted very well by Planck's law predictions at 195 and 283 °C (solid). However, for λ < 8 μm, the measured values are larger than the Planck's law predictions [inset of (c)]. (e) Ratios of thermal-emission spectra between experiment and Planck's law for the fused-silica window: green line for 200 °C (experiment at 200 °C, Planck's law at 195 °C) and red line for 300 °C (experiment at 300 °C, Planck's law at 283 °C). (f) The same as (e) but for the CNT blackbody. The rapid increase of ratios for fused silica below 5 μm is due to the corresponding increase of transmission [inset of (e)]. Ratios from the corresponding full model calculations are also plotted in (e) using dotted black lines.

One way to determine the actual surface temperatures of the samples is to match the experimental spectra in the wavelength range where the samples are opaque to the simple theoretical predictions (i.e., using Planck's law and the sample's emissivity), using temperature as a fitting parameter. As shown in Fig. 3(d), the measured CNT blackbody spectra for stage temperatures of 300 and 200 °C can be well fitted by Planck's law at 282 and 193 °C, respectively. The CNTs have sufficiently high optical loss that the measured thermal emission only comes from the surface [30], implying that these are the true surface temperatures. A similar analysis of the fused-silica window for λ > 8 μm, where it is highly opaque [24], yields surface temperatures of 283 and 195 °C [Fig. 3(c)]. The results agreed well with infrared-camera images analyzed using commercial thermography software (Sec. S3, Supplementary Information).

For $\lambda < 8$ μm, the measured spectra of fused silica are slightly larger than Planck's law predictions [inset, Fig. 3(c)]. In the spectral region between 5 and 8 μm, the penetration depth is on the order of hundreds of micrometers [24], comparable to the sample thickness, resulting in measured thermal emission from the volume of the sample. The contributions from the hotter regions beneath the surface led to more emission compared to the theoretical prediction that assumed a uniform temperature equal to that of the top surface. Note that to clearly see this effect, one has to zoom into the measured data [inset of Fig. 3(c)]. To better demonstrate this phenomenon and show that this effect is larger than the measurement noise or other errors, the ratios of the experimental spectra to the Planck's law are plotted in Figs. 3(e) and (f) for the fused-silica window and the CNT blackbody, respectively.

The CNT blackbody ratio is close to unity across the entire spectral range of the measurement (4.5 to 16 μm). For the fused-silica window, this ratio is close to unity only for $\lambda > 10$ μm, but is larger for $\lambda < 8$ μm, a clear indication of the contribution of thermal emission from the hotter regions beneath the surface. The fluctuations of the ratio near 9 μm are due to the temperature dependence of the vibrational resonances of silica [31], resulting in relatively strong temperature dependence of the optical properties in this wavelength range (Sec. S4, Supplementary Information). In Fig. 3(e), the difference between the ratio and unity for $\lambda < 8$ μm at 300 °C is about three times larger than that for 200 °C, in agreement with the relative magnitude of the corresponding temperature gradient: a 17 °C drop across the window at 300 °C vs. a 5 °C drop at 200 °C .

## Theoretical modeling and temperature extraction

To retrieve the temperature profile beneath the surface from the thermal-emission measurements, we modeled the 1-mm-thick fused-silica window as a discretized thin-film stack, with each layer having a different temperature, $T_j = T(z)$. Here, $j$ represents the layer number of the discretized stack along the depth direction $z$ [Fig. 4(a)]. Our model can handle the case for layers with different material properties [32], but we assumed that the optical properties [$n(\lambda)$ and $\kappa(\lambda)$] are temperature-independent within the small temperature interval in our experiment (e.g., interval of 17 °C when the stage is at 300 °C).

The thermal emission from each layer $I_j$ can be calculated using the fluctuation–dissipation theorem and the dyadic Green's function [33], [34]:

$$I_j = \bar{\epsilon}_j(\lambda) I_{BB}(\lambda, T_j) \tag{3}$$

In Eq. 3, we define a special variable $\bar{\epsilon}_j(\lambda)$, which we termed *local emissivity*, that quantifies the portion of thermal emission from layer $j$ that reaches free space. This value depends on not only the optical properties of layer $j$, but also the optical properties (and, more generally, the overall dielectric environment) of each remaining layer surrounding it. The concept of $\bar{\epsilon}_j(\lambda)$ has been proposed previously [34] and can be calculated via the scattering matrix method [33]. Summing the contributions from all layers leads to the total emitted spectrum:

$$I(\lambda) = \sum_{j=1}^{N} \bar{\epsilon}_j(\lambda) I_{BB}(\lambda, T_j) \tag{4}$$

Note that Eq. (4) can be generalized to the continuous case as:

$$I(\lambda) = \int_{all\ z} \bar{\epsilon}(\lambda, z) I_{BB}[\lambda, T(z)] dz, \tag{5}$$

where $\bar{\epsilon}(\lambda, z)$ is the *local emissivity density* and is related to $\bar{\epsilon}_j(\lambda)$ via $\bar{\epsilon}_j(\lambda) = \int_{z_1}^{z_2} \bar{\epsilon}(\lambda, z) dz$, where the integration covers the region of layer $j$.

In one-dimensional heat transfer and considering only thermal conductance, the heat-energy flux density is the product of thermal conductivity ($K$) and the temperature gradient ($\partial T/\partial z$): $q = -K\ \partial T/\partial z$, according to Fourier's law of heat conduction [35]. Note that here we do not consider the impact of radiative heat transfer because it is small enough to be considered negligible compared to thermal conductance, though it can contribute to the temperature distribution for thicker samples and/or higher temperatures [36]. In steady state, $q$ is constant across the window (from the bottom to the top surface), and therefore $\partial T/\partial z$ is also constant because $K$ is also very close to constant; *i.e.*, the temperature inside the window is a linear function of depth. Thus, we assumed a linearly decreasing temperature from the bottom (300 °C) to the top (283 °C) for the 1-mm-thick fused-silica window. The optical properties ($n$ and $\kappa$) of the fused silica were measured using variable-angle spectroscopic ellipsometry with the sample heated to 300 °C (Sec. S4, Supplementary Information). Using Eq. 4 and the expected linear temperature distribution, we calculated the emission spectrum, plotted in Fig. 4(b), which is in excellent agreement with the experimental spectrum.

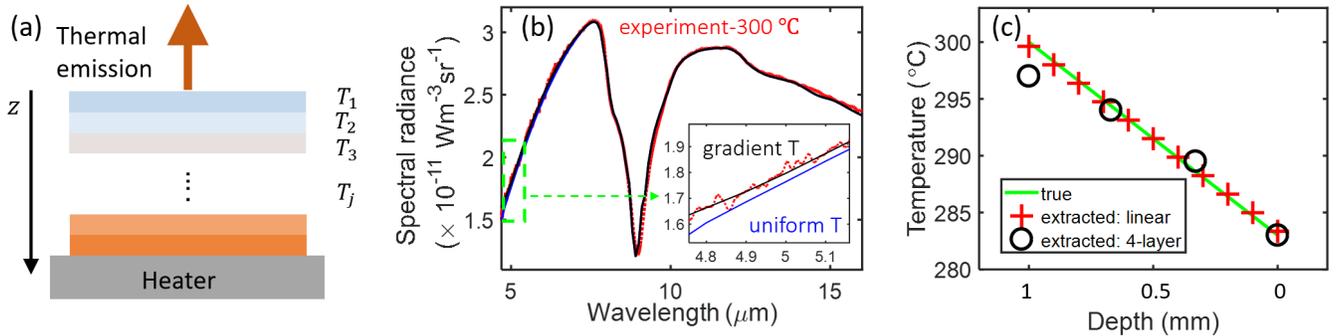

Figure 4. (a) The 1-mm-thick fused-silica window is modeled as a multilayer system, with each layer having the same refractive indices, but different temperatures $T_j$. The total emission spectrum is obtained by summing contributions from all individual layers. (b) The calculated thermal-emission spectrum (black, assuming a gradient temperature) for a 1-mm-thick fused-silica window on top of a 300 °C heater agrees well with the experimental spectrum (red). The blue line shows the calculated spectrum from the same window but assuming a uniform temperature of 283 °C. (c) Extracted temperature profiles from the experimental spectrum shown in (b) using different approaches: red crosses assuming a linear temperature profile, and black circles for a 4-layer model without any assumptions about the shape of the temperature distribution. Both extracted profiles match well with the actual temperature profile (green).

To further compare the model with the experiment, we performed a calculation assuming a uniform temperature of 283 °C across the same fused-silica window [Fig. 4(b)]. The two calculated spectra overlap perfectly in the opaque region ($\lambda > 8$ μm) but differ in the semitransparent region ($\lambda < 8$ μm). The ratio of these two model calculations is plotted in Fig. 3(e), in agreement with the experiment.

The one-to-one relationship between the temperature and the spectrum in Eq. 4 enables the extraction of the depth-dependent temperature distribution from the thermal-emission spectrum, assuming known optical properties, *i.e.*, the local emissivity $\bar{\epsilon}_j(\lambda)$. More specifically, for the case of an $N$-layer structure, the $N$ unknown temperatures $T_j, j = 1,2,..,N$ can in principle be determined by solving a set of at least $N$ equations obtained by picking data at $N$ independent wavelength from Eq. 4. We demonstrated this process for an ideal thermal-emission spectrum in Sec. 5.1, Supplementary Information. In real experiments, however, the measured spectra are inevitably noisy, which complicates the inversion process (see more details in Sec. 5.2, Supplementary Information).

Temperature extraction from noisy measurements has been investigated in the microwave community, and several inversion methods have been proposed [19], [21], [36], [37]. One standard approach is to add constraints to the temperature distribution when some additional information is available. For example, the temperature inside certain biological subjects has been extracted with the temperature distribution assumed to be exponential [38], [39]. In the present experiment, the temperature inside the silica window is expected to be a linear function of depth. This constraint makes temperature extraction from the experimental spectrum quite straightforward [red crosses, Fig. 4(c)]. We have also numerically demonstrated that this approach can be applied to other temperature distributions, such as a hot layer embedded in the volume of a material (see more details in Sec. 5.2, Supplementary Information).

In some cases, the shape of the temperature distribution may be completely unknown. If few or no constraints are known, spatial resolution (i.e., the number of independent temperatures as a function of depth) must be sacrificed for more-robust temperature extraction (see more details in Sec. 5.2, Supplementary Information). We demonstrated this with our experimental spectrum from Fig. 4(b), in which the 1-mm-thick fused-silica window was modeled as a four-layer system with four unknown temperatures as a function of depth. The temperature of each layer was then determined from the combination that minimized the least squared error:

$$\sum_{\lambda_1}^{\lambda_2}\{I_{exp}(\lambda) - \sum_{j=1}^{4}\bar{\epsilon}_j(\lambda)I_{BB}(\lambda, T_j)\}^2 \tag{6}$$

The extracted temperature profile [black circles, Fig. 4(c)] agrees well with the expected temperature profile.

## Discussion

We introduced and demonstrated *depth thermography*: a technique that can remotely resolve temperature information beneath the surface of an infrared-semitransparent object by analyzing the spectrum of thermal emission emerging from the surface. Depth thermography may compete with or be complementary to multi-frequency microwave radiometry [17]–[21], the other method to remotely and nondestructively measure temperature distributions as a function of depth. The preferred method(s) will depend in part on the regions of transparency/semitransparency of the material of interest. For example, Earth's atmosphere has transmission bands in the visible, mid-infrared, and microwave ranges [21], [40], and it may be beneficial to simultaneously perform measurements using more than one semitransparent region adjacent to these transparency windows.

It is also instructive to directly compare infrared depth thermography with microwave radiometry by analyzing thermal-emission measurements at wavelengths of, e.g., 5 μm and 5 cm. First, the size of the minimum resolvable spatial features using these techniques is bounded by the diffraction limit, and thus infrared depth thermography can, in principle, resolve microscopic features of ~10 microns — a 10,000-fold improvement over the microwave method. Next, the sensitivity in resolving depth-temperature variation is proportional to $\partial L_{BB}(\lambda, T)/\partial T$, which increases as the wavelength becomes shorter, resulting in about a 10× improvement moving from the microwave to the infrared. Finally, close to room temperature, there is much more signal in the infrared vs. the microwave: according to Planck's law (Eq. 3), $I_{BB}(\lambda = 5 \text{ μm}, T = 300 \text{ K}) \sim 10^{13} I_{BB}(\lambda = 5 \text{ cm}, T = 300 \text{ K})$, though this does not indicate that depth thermography is 10 trillion times better than microwave radiometry in terms of the signal-to-noise ratio (SNR), which also depends on the detector performance, the detection volume and solid angle, and the overall collection efficiency of the measurement system. Note that for measurements with higher spatial resolution, the total emitted power per pixel decreases because the emitted power is proportional to the emitting area. Thus, increasing the resolution from centimeters to microns decreases the emitting power by about $10^8$, and the $10^{13}$ increase in per-area signal in the infrared vs. the microwave region makes it possible to perform diffraction-limited measurements in the infrared.

Because depth thermography is a volumetric temperature-measurement technique, it can be combined with heat-transfer models to perform noncontact (remote) measurements of the thermal conductivity of materials. In particular, depth thermography can enable measurements of cross-plane thermal conductivity of thin-film materials, which can be challenging with conventional methods [41]. For example, the temperature extraction shown in Eq. 6 and Fig. 4 is effectively an all-optical measurement of the thermal conductivity of fused silica. While, as presented in this work, depth thermography enables the measurement of only the vertical (cross-plane) component of the thermal conductivity, it can be extended to measurements of the entire thermal-conductivity tensor by performing 3D measurements (using 2D scanning or a hyperspectral infrared camera) of a sample that was heated using a localized heater in one particular spot. This can be extended even further to volumes where heat transfer is a combination of multiple heat-transfer mechanisms (conduction, convention, and/or thermal radiation), which is of relevance to high-temperature and/or high-pressure gasses [42] as well as high-temperature ionic liquids [43] and molten salts [44].

As described in this paper, depth thermography enables the extraction of temperature distributions or thermal conductivities given known optical properties. However, the precise optical properties of a sample may also be unknown. We anticipate that depth thermography can be extended to simultaneous measurements of the optical and thermal properties by measuring thermal-emission spectra at different angles and polarizations, similar to variable-angle spectroscopic ellipsometry [45], or by integrating thermal-emission spectroscopy with conventional methods of materials characterization in the same instrument.

## Materials and Methods

The 1-mm-thick fused-silica window was purchased from Thorlabs. The laboratory blackbody reference (~500-μm-tall vertically-aligned carbon nanotube forest grown on silicon substrate) was purchased from NanoTechLabs, Inc. The thermal emission spectra were measured using an FTS (Bruker VERTEX 70) [28]. The thermal-emission signal from the sample was collected by a parabolic mirror (with a numerical aperture of ~0.05), sent into a moving-mirror Michelson interferometer, and then detected by a liquid-nitrogen-cooled mercury-cadmium-telluride (MCT) detector. In our setup, the beam path between the interferometer and detector includes several mirrors and apertures. The infrared imaging was carried out using a FLIR A325sc camera, sensitive to the 7.5 to 13 $\mu$m range with commercial software from FLIR. The ellipsometry was performed using an ellipsometer from J.A. Woollam Co., with incident angles of 35, 45 and 55°, and the data were analyzed using WVASE software.

In calibrating the FTS measurement system, two known references (fused-silica and sapphire) were used to find the sample-independent and sample-dependent background. The system response was determined from the thermal-emission signal measured from a reference at two different temperatures. Details of the calibration are shown in Sec. 1 of the Supplemental Information.

In the model calculation, we divided the fused-silica window into 11 layers and first calculated the local emissivity of each individual layer using the scattering-matrix method. The total thermal emission was then obtained by summing the product of the local contribution from all 11 layers. In extracting the temperature distribution, the nonlinear equation solver "lsqnonlin" in Matlab was used to find the solution within certain bounded regions. Details of temperature extraction can be found in Sec. 5 of the Supplemental Information.

## Acknowledgments

This work is supported by the Office of Naval Research (N00014-16-1-2556) and the Department of Energy NEUP (DE-NE0008680). The authors thank Tom Beechem from Sandia National Laboratory and Scott Sanders from UW-Madison for helpful discussions. MK also thanks T. Z. Choy-Kats for support.

# References

[1]  D. G. Baranov, Y. Xiao, I. A. Nechepurenko, A. Krasnok, A. Alù, and M. A. Kats, "Nanophotonic engineering of far-field thermal emitters," *Nat. Mater.*, vol. 18, pp. 920–930, May 2019.

[2]  M. Fuchs and C. B. Tanner, "Infrared Thermometry of Vegetation1," *Agron. J.*, vol. 58, no. 6, p. 597, 1966.

[3]  G. Gaussorgues and S. Chomet, *Infrared Thermography*. Springer Netherlands, 1994.

[4]  C. Meola and G. M. Carlomagno, "Recent advances in the use of infrared thermography," *Meas. Sci. Technol.*, vol. 15, no. 9, pp. 27–58, Sep. 2004.

[5]  J. A. Voogt and T. R. Oke, "Thermal remote sensing of urban climates," *Remote Sens. Environ.*, vol. 86, no. 3, pp. 370–384, Aug. 2003.

[6]  E. Gartenberg and A. S. Oberts, "Twenty-five years of aerodynamic research with infrared imaging," *J. Aircr.*, vol. 29, no. 2, pp. 161–171, Mar. 1992.

[7]  R. D. Hudson and J. W. Hudson, "The military applications of remote sensing by infrared," *Proc. IEEE*, vol. 63, no. 1, pp. 104–128, 1975.

[8]  B. B. Lahiri, S. Bagavathiappan, T. Jayakumar, and J. Philip, "Medical applications of infrared thermography: A review," *Infrared Phys. Technol.*, vol. 55, no. 4, pp. 221–235, Jul. 2012.

[9]  S. Bagavathiappan, B. B. Lahiri, T. Saravanan, J. Philip, and T. Jayakumar, "Infrared thermography for condition monitoring – A review," *Infrared Phys. Technol.*, vol. 60, pp. 35–55, Sep. 2013.

[10] W. J. Wright, R. B. Schwarz, and W. D. Nix, "Localized heating during serrated plastic flow in bulk metallic glasses," *Mater. Sci. Eng. A*, vol. 319–321, pp. 229–232, Dec. 2001.

[11] M. Arik, C. A. Becker, S. E. Weaver, and J. Petroski, "Thermal management of LEDs: package to system," in *Third international conference on solid state lighting*, 2004, vol. 5187, p. 64.

[12] V. Spagnolo *et al.*, "Temperature profile of GaInAs/AlInAs/InP quantum cascade-laser facets measured by microprobe photoluminescence," *Appl. Phys. Lett.*, vol. 78, no. 15, pp. 2095–2097, Apr. 2001.

[13] C. A. Evans, V. D. Jovanovic, D. Indjin, Z. Ikonic, and P. Harrison, "Investigation of Thermal Effects in Quantum-Cascade Lasers," *IEEE J. Quantum Electron.*, vol. 42, no. 9, pp. 857–865, Sep. 2006.

[14] A. Jain, R. E. Jones, Ritwik Chatterjee, S. Pozder, and Zhihong Huang, "Thermal modeling and design of 3D integrated circuits," in *11th Intersociety Conference on Thermal and Thermomechanical Phenomena in Electronic Systems*, 2008, pp. 1139–1145.

[15] J. Serp *et al.*, "The molten salt reactor (MSR) in generation IV: Overview and perspectives," *Prog. Nucl. Energy*, vol. 77, pp. 308–319, Nov. 2014.

[16] C. Taylor, *The internal combustion engine in theory and practice*. Cambridge Mass.: MIT Press, 1968.

[17] A. H. Barrett and P. C. Myers, "Subcutaneous temperatures: a method of noninvasive sensing.," *Science*, vol. 190, no. 4215, pp. 669–71, Nov. 1975.

[18] F. Bardati, V. J. Brown, and G. Di Bernardo, "Multi-frequency Microwave Radiometry for Retrieval of Temperature Distributions in the Human Neck," *J. Photogr. Sci.*, vol. 39, no. 4, pp. 157–160, Jul. 1991.

[19] S. Mizushina, T. Shimizu, K. Suzuki, M. Kinomura, H. Ohba, and T. Sugiura, "Retrieval of Temperature-Depth Profiles in Biological Objects from Multi-Frequency Microwave Radiometric Data," *J. Electromagn. Waves Appl.*, vol. 7, no. 11, pp. 1515–1548, Jan. 1993.

[20] J. W. Hand *et al.*, "Monitoring of deep brain temperature in infants using multi-frequency microwave radiometry and thermal modelling," *Phys. Med. Biol.*, vol. 46, no. 7, pp. 1885–1903, Jul. 2001.

[21] F. Solheim *et al.*, "Radiometric profiling of temperature, water vapor and cloud liquid water using various inversion methods," *Radio Sci.*, vol. 33, no. 2, pp. 393–404, Mar. 1998.

[22] M. Planck, "Ueber das Gesetz der Energieverteilung im Normalspectrum," *Ann. Phys.*, vol. 309, no. 3, pp. 553–563, 1901.

[23] G. Kirchhoff, "Über das Verhältnis zwischen dem Emissionsvermögen und dem Absorptionsvermögen der Körper für Wärme und Licht," in *Von Kirchhoff bis Planck*, Wiesbaden: Vieweg+Teubner Verlag, 1978, pp. 131–151.

[24] R. Kitamura, L. Pilon, and M. Jonasz, "Optical constants of silica glass from extreme ultraviolet to far infrared at near room temperature," *Appl. Opt.*, vol. 46, no. 33, p. 8118, Nov. 2007.

[25] W. Kaiser, W. G. Spitzer, R. H. Kaiser, and L. E. Howarth, "Infrared Properties of CaF2, SrF2, and BaF2," *Phys. Rev.*, vol. 127, no. 6, pp. 1950–1954, Sep. 1962.

[26] S. M. Rytov, Y. A. Kravtsov, and V. I. Tatarskii, *Priniciples of Statistical Radiophysics*. Springer, 1989.

[27] E. H. Ratcliffe, "Thermal conductivities of fused and crystalline quartz," *Br. J. Appl. Phys.*, vol. 10, no. 1, pp. 22–25, Jan. 1959.

[28] Y. Xiao *et al.*, "Measuring Thermal Emission Near Room Temperature Using Fourier-Transform Infrared Spectroscopy," *Phys. Rev. Appl.*, vol. 11, no. 1, p. 014026, Jan. 2019.

[29] K. Mizuno *et al.*, "A black body absorber from vertically aligned single-walled carbon nanotubes.," *Proc. Natl. Acad. Sci. U. S. A.*, vol. 106, no. 15, pp. 6044–7, Apr. 2009.

[30] H. Ye, X. J. Wang, W. Lin, C. P. Wong, and Z. M. Zhang, "Infrared absorption coefficients of vertically aligned carbon nanotube films," *Appl. Phys. Lett.*, vol. 101, no. 14, p. 141909, Oct. 2012.

[31] J. Kischkat *et al.*, "Mid-infrared optical properties of thin films of aluminum oxide, titanium dioxide, silicon dioxide, aluminum nitride, and silicon nitride," *Appl. Opt.*, vol. 51, no. 28, p. 6789, Oct. 2012.

[32] Y. Xiao, N. A. Charipar, J. Salman, A. Piqué, and M. A. Kats, "Nanosecond mid-infrared pulse generation via modulated thermal emissivity," *Light Sci. Appl.*, vol. 8, no. 1, p. 51, Dec. 2019.

[33] M. Francoeur, M. Pinar Mengüç, and R. Vaillon, "Solution of near-field thermal radiation in one-dimensional layered media using dyadic Green's functions and the scattering matrix method," *J. Quant. Spectrosc. Radiat. Transf.*, vol. 110, no. 18, pp. 2002–2018, 2009.

[34] L. P. Wang, S. Basu, and Z. M. Zhang, "Direct and Indirect Methods for Calculating Thermal Emission From Layered Structures With Nonuniform Temperatures," *J. Heat Transfer*, vol. 133, no. 7, p. 072701, Jul. 2011.

[35] M. N. Özışık, *Heat conduction*. Wiley, 1993.

[36] F. Bardati, M. Mongiardo, and D. Solimini, "Inversion of Microwave Thermographic Data by the Singular Function Method," in *MTT-S International Microwave Symposium Digest*, vol. 85, pp. 75–77.

[37] E. . Westwater, "Ground-Based Passive Probing Using the Microwave Spectrum of Oxygen," *J. Res.*, vol. 69, no. 9, pp. 1201–1211, 1965.

[38] V. S. Troitskii *et al.*, "Measuring the temperature depth profile in biological subjects through characteristic thermal microwave emissions," *Radiophys. Quantum Electron.*, vol. 29, no. 1, pp. 52–57, Jan. 1986.

[39] S. Mizushina, T. Shimizu, K. Suzuki, and T. Sugiura, "A Temperature Profile Retrieval Technique from

[26] S. M. Rytov, Y. A. Kravtsov, and V. I. Tatarskii, *Priniciples of Statistical Radiophysics*. Springer, 1989.

[27] E. H. Ratcliffe, "Thermal conductivities of fused and crystalline quartz," *Br. J. Appl. Phys.*, vol. 10, no. 1, pp. 22–25, Jan. 1959.

[28] Y. Xiao *et al.*, "Measuring Thermal Emission Near Room Temperature Using Fourier-Transform Infrared Spectroscopy," *Phys. Rev. Appl.*, vol. 11, no. 1, p. 014026, Jan. 2019.

[29] K. Mizuno *et al.*, "A black body absorber from vertically aligned single-walled carbon nanotubes.," *Proc. Natl. Acad. Sci. U. S. A.*, vol. 106, no. 15, pp. 6044–7, Apr. 2009.

[30] H. Ye, X. J. Wang, W. Lin, C. P. Wong, and Z. M. Zhang, "Infrared absorption coefficients of vertically aligned carbon nanotube films," *Appl. Phys. Lett.*, vol. 101, no. 14, p. 141909, Oct. 2012.

[31] J. Kischkat *et al.*, "Mid-infrared optical properties of thin films of aluminum oxide, titanium dioxide, silicon dioxide, aluminum nitride, and silicon nitride," *Appl. Opt.*, vol. 51, no. 28, p. 6789, Oct. 2012.

[32] Y. Xiao, N. A. Charipar, J. Salman, A. Piqué, and M. A. Kats, "Nanosecond mid-infrared pulse generation via modulated thermal emissivity," *Light Sci. Appl.*, vol. 8, no. 1, p. 51, Dec. 2019.

[33] M. Francoeur, M. Pinar Mengüç, and R. Vaillon, "Solution of near-field thermal radiation in one-dimensional layered media using dyadic Green's functions and the scattering matrix method," *J. Quant. Spectrosc. Radiat. Transf.*, vol. 110, no. 18, pp. 2002–2018, 2009.

[34] L. P. Wang, S. Basu, and Z. M. Zhang, "Direct and Indirect Methods for Calculating Thermal Emission From Layered Structures With Nonuniform Temperatures," *J. Heat Transfer*, vol. 133, no. 7, p. 072701, Jul. 2011.

[35] M. N. Özışık, *Heat conduction*. Wiley, 1993.

[36] F. Bardati, M. Mongiardo, and D. Solimini, "Inversion of Microwave Thermographic Data by the Singular Function Method," in *MTT-S International Microwave Symposium Digest*, vol. 85, pp. 75–77.

[37] E. . Westwater, "Ground-Based Passive Probing Using the Microwave Spectrum of Oxygen," *J. Res.*, vol. 69, no. 9, pp. 1201–1211, 1965.

[38] V. S. Troitskii *et al.*, "Measuring the temperature depth profile in biological subjects through characteristic thermal microwave emissions," *Radiophys. Quantum Electron.*, vol. 29, no. 1, pp. 52–57, Jan. 1986.

[39] S. Mizushina, T. Shimizu, K. Suzuki, and T. Sugiura, "A Temperature Profile Retrieval Technique from

Multi-Frequency Microwave Radiometric Data for Medical Applications," in *AMPC Asia-Pacific Microwave Conference,* 1992, vol. 1, pp. 233–236.

[40]   J. W. Salisbury and D. M. D'Aria, "Emissivity of terrestrial materials in the 8–14 μm atmospheric window," *Remote Sens. Environ.*, vol. 42, no. 2, pp. 83–106, Nov. 1992.

[41]   D. Zhao, X. Qian, X. Gu, S. A. Jajja, and R. Yang, "Measurement Techniques for Thermal Conductivity and Interfacial Thermal Conductance of Bulk and Thin Film Materials," *J. Electron. Packag.*, vol. 138, no. 4, p. 040802, Oct. 2016.

[42]   S. T. Sanders, J. Wang, J. B. Jeffries, and R. K. Hanson, "Diode-laser absorption sensor for line-of-sight gas temperature distributions," *Appl. Opt.*, vol. 40, no. 24, p. 4404, Aug. 2001.

[43]   M. E. V. Valkenburg, R. L. Vaughn, M. Williams, and J. S. Wilkes, "Thermochemistry of ionic liquid heat-transfer fluids," *Thermochim. Acta*, vol. 425, no. 1–2, pp. 181–188, Jan. 2005.

[44]   E. S. Chaleff, T. Blue, and P. Sabharwall, "Radiation Heat Transfer in the Molten Salt FLiNaK," *Nucl. Technol.*, vol. 196, no. 1, pp. 53–60, Oct. 2016.

[45]   R. M. A. Azzam and N. M. Bashara, *Ellipsometry and polarized light*. Amsterdam: North-Holland, 1987.

[46]   R. Boyd, *Radiometry and the detection of optical radiation*. John Wiley & Sons, 1983.

# Depth thermography: non-invasive 3D temperature profiling using infrared thermal emission

**Yuzhe Xiao[1], Chenghao Wan[1,2], Alireza Shahsafi[1], Jad Salman[1], and Mikhail A. Kats[1,2,3,*]**

[1]*Department of Electrical and Computer Engineering, University of Wisconsin-Madison, Madison, Wisconsin 53706, USA*
[2]*Materials Science and Engineering, University of Wisconsin-Madison, Madison, Wisconsin 53706, USA*
[3]*Department of Physics, University of Wisconsin-Madison, Madison, Wisconsin, USA 53706*
*Email address:* mkats@wisc.edu

## S1: Fourier-transform spectrometer (FTS) calibration

The measured thermal-emission signal is inevitably mixed with a background because every component of the FTS measurement system emits thermal radiation. The background could come from the instrument and the surrounding environment and can be sample-dependent. For example, there could be background emission from the room where the instrument is located that is reflected or scattered into the FTS beam path. In general, the measured signal from an emitter $x$ can be expressed by the following expression:

$$S_x(\lambda) = m(\lambda)[I_x(\lambda) + B_x(\lambda)], \tag{S1}$$

where $I_x(\lambda)$ is the true emission spectrum from the emitter, $B_x(\lambda)$ is the background emission, and $m(\lambda)$ is the system response, including the collection efficiency of the setup and the responsivity of the detector. A non-scattering opaque emitter can only affect the background through reflection. In this case, we can write $B_x(\lambda) = R_x(\lambda, T)B_1(\lambda) + B_2(\lambda)$, where $R_x(\lambda, T)B_1(\lambda)$ and $B_2(\lambda)$ represent the sample-dependent and sample-independent contribution of the background emission, respectively. Therefore, Eq. S1 becomes the following for an opaque and non-scattering emitter:

$$S_x(\lambda) = m(\lambda)[I_x(\lambda) + R_x(\lambda, T)B_1(\lambda) + B_2(\lambda)]. \tag{S2}$$

If all parts of the emitter that we are measuring are in thermal equilibrium (i.e., the entire emitter has a single uniform temperature), then the emission spectrum from the emitter can be written as: $I_x(\lambda) = \epsilon_x(\lambda, T)I_{BB}(\lambda, T)$, where $\epsilon_x(\lambda, T)$ is the emissivity and $I_{BB}(\lambda, T)$ is the blackbody spectrum at sample temperature $T$ given by Planck's law. In this case, Eq. S2 becomes:

$$S_x(\lambda, T) = m(\lambda)[\epsilon_x(\lambda)I_{BB}(\lambda, T) + R_x(\lambda, T)B_1(\lambda) + B_2(\lambda)]. \tag{S3}$$

Looking at Eq. S2, to extract the true emission signal $I_x(\lambda)$ from the measured data $S_x(\lambda)$, one needs to determine

$m(\lambda)$, $B_1(\lambda)$, $B_2(\lambda)$ and $R_x(\lambda)$. The first three terms are related to the FTS and can be determined by measuring two known non-scattering, opaque references $\alpha$ and $\beta$, whose emissivity and reflection coefficient do not change significantly with temperature, as follows:

#1: The system response function can be obtained from thermal emission measured from reference $\alpha$ at two different temperatures $T_1$ and $T_2$:

$$m(\lambda) = \frac{S_\alpha(\lambda, T_1) - S_\alpha(\lambda, T_2)}{\epsilon_\alpha(\lambda)[I_{BB}(\lambda, T_1) - I_{BB}(\lambda, T_2)]} \tag{S4}$$

#2: With the known system response function $m(\lambda)$, the total backgrounds for $\alpha$ and $\beta$ can be obtained from thermal emission measured at temperature $T_1$ (or $T_2$):

$$B_\alpha(\lambda) = \frac{S_\alpha(\lambda, T_1)}{m(\lambda)} - \epsilon_\alpha(\lambda) I_{BB}(\lambda, T_1) \tag{S5}$$

$$B_\beta(\lambda) = \frac{S_\beta(\lambda, T_1)}{m(\lambda)} - \epsilon_\beta(\lambda) I_{BB}(\lambda, T_1) \tag{S6}$$

#3: $B_1(\lambda)$ and $B_2(\lambda)$ can then be determined from $B_\alpha(\lambda)$ and $B_\beta(\lambda)$:

$$B_1(\lambda) = \frac{B_\alpha(\lambda) - B_\beta(\lambda)}{R_\alpha(\lambda) - R_\beta(\lambda)} \tag{S7}$$

$$B_2(\lambda) = B_\alpha(\lambda) - R_\alpha(\lambda) B_1(\lambda) \tag{S8}$$

We measured thermal emission from polished wafers of fused silica and sapphire—known references—to calibrate our measurements. We characterized the two references by measuring their emissivity $\epsilon(\lambda)$ by taking the difference of emission at two temperatures (60 and 100 °C, where the emissivity of both fused silica and sapphire does not change) and normalizing to that of the blackbody reference:

$$\epsilon(\lambda) = \epsilon_{BB}(\lambda) \frac{S(\lambda, T_1) - S(\lambda, T_2)}{S_{BB}(\lambda, T_1) - S_{BB}(\lambda, T_2)} \tag{S9}$$

Their reflectances are obtained as $R(\lambda) = 1 - \epsilon(\lambda)$ according to Kirchhoff's law [S1]. Note here at these low temperatures, the temperature gradient can be safely neglected. Figure S1(a) shows the measured raw (but Fourier-transformed) thermal-emission signal from wafers of sapphire and fused silica at 50 and 75 °C. In the measurement, the samples were tilted by 10° to avoid multiple reflections between the sample and the interferometer [S2]. The calibrated system response $m(\lambda)$ and the backgrounds $B_1(\lambda)$ and $B_2(\lambda)$ are plotted in Fig. S1 (b, c). Note that our calculated $B_2$ is negative because in our setup the constant background emission originates from components after the interferometer [S2]. Also note that the artificial peaks in $B_1(\lambda)$ and $B_2(\lambda)$ near 7 and 11 $\mu$m come from the fact that the reflection of sapphire and fused silica are very close to each other near these two wavelengths, leading to sharp features. The calibrated values of $B_1(\lambda)$ and $-B_2(\lambda)$ can be well represented by thermal emission from room-temperature emitters with effective emissivity values of 0.95 for $B_1$ and 0.94 for $B_2$, as shown by the dotted black curves in Fig. S1(c). The effective emissivity of $B_1$ (0.95) is expected because the sample is placed inside a sample compartment in our FTS; the enclosure of the sample compartment can be approximately treated as a blackbody with emissivity close to unity [S3]. The effective emissivity of $B_2$ (0.94) indicates that there is a significant amount of background after the interferometer in our FTS [S2].

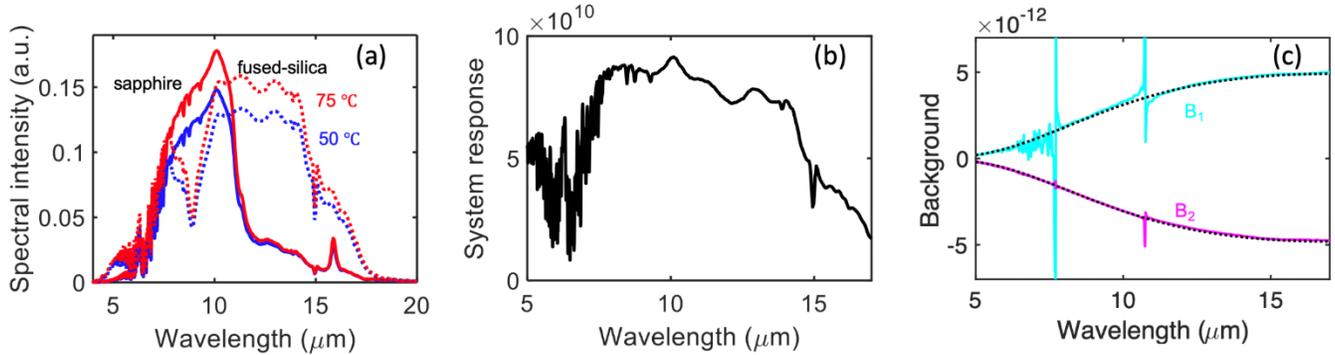

Figure S1. (a): Measured thermal emission from polished wafers of sapphire (solid) and fused silica (dotted) at 50 (blue) and 75 °C (red). (b): Calibrated system response $m(\lambda)$ of our FTS. (c): Calibrated $B_1$ (purple) and $B_2$ (cyan) of our measurement system. Dotted lines show the corresponding fitting of $B_1$ and $B_2$.

## S2. Reflectance measurement of the fused-silica window

As shown in Eq. S2, the reflection coefficient of the fused-silica window needs to be determined to obtain the true thermal-emission signal. The reflectance of the 1-mm fused silica was measured with our FTS using a reflective microscope objective with NA = 0.4. Figure S2 shows the measured reflectance at room-temperature, 200 and 300 °C. As shown here, the reflectance of fused silica changes with temperature [S4]. The peak between 8 and 10 $\mu m$ decreases as temperature increases, but the change in reflectance (and, we assume, all optical properties) is negligible away from the vibrational resonances.

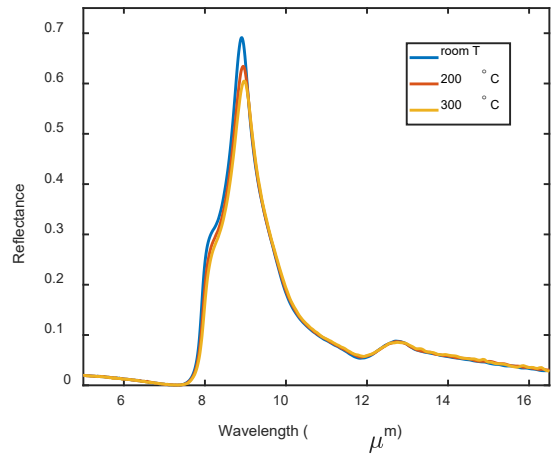

Figure S2. Measured temperature-dependent reflectance of the 1-mm-thick fused silica.

Note that our microscope reflectance measurements were performed with an objective with NA = 0.4 centered around the normal, whereas our emission setup had NA = 0.05 centered around an angle of 10°. Despite these differences, the measured reflectance using the microscope setup is very close to the reflectance of the fused-silica window in the emission setup. This is because the averaged p- and s-polarized reflectance does not change within a small range of angles near the normal direction. To demonstrate this, we calculated the expected reflectance of 1-mm-thick fused silica using the optical properties extracted from spectroscopic ellipsometry measurement at 300 ℃ (Fig. S5). The polarization-averaged reflectance does not change appreciably with incident angle for angles less than 25°.

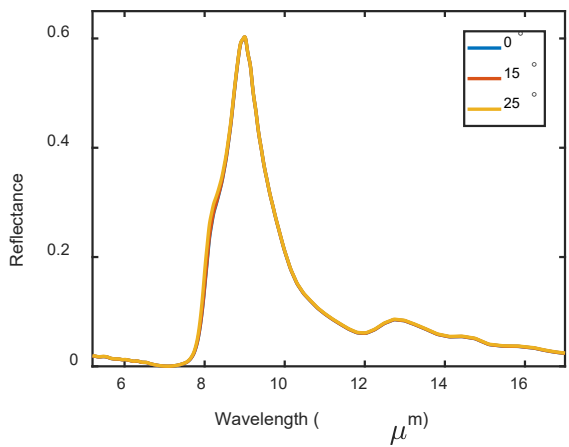

Figure S3. Calculated reflectance averaged over p- and s-polarization for the 1-mm-thick fused-silica window using the optical properties extracted from spectroscopic ellipsometry at 300 ℃.

### S3: Measuring surface temperature with a commercial infrared camera

To double check the surface temperature of the samples obtained by fitting the measured thermal-emission spectra, we used a mid-infrared camera (FLIR A325sc with software from FLIR). Our infrared camera has a bandwidth from 7.5 to 13 μm, where both the fused-silica window and the CNT blackbody reference are opaque [S5]. The infrared-camera software returns a map of temperature once a wavelength-integrated emissivity value is assigned in the camera software (we refer to this as $\epsilon_{set}$).

To measure the surface temperature of both samples when heated by a 300 °C heater, we first heated both samples to 50 °C, putting the samples in firm contact with the heater stage. At 50 °C, the temperature gradient between the top and the bottom of these samples can be safely neglected, so we adjusted $\epsilon_{set}$ such that the camera reading returns 50 °C. We found $\epsilon_{set}$ of 0.89 for fused silica and 0.97 for the CNT blackbody led the camera software to return 50 °C [Figure S4(a) and (c)]. Then both samples were further heated up by setting the heater temperature to 300 °C. The corresponding temperature readings from the camera are shown in Figure S4(b) and (d). As shown here, the measurements from the infrared camera (surface temperature of 283.0 °C for the fused-silica window and 282.3 °C for the CNT blackbody) agree quite well with the values obtained by fitting the emission spectrum (surface temperature of 283 °C for fused silica and 282 °C for CNT blackbody, Fig. 3).

Note that the infrared-camera software also has several assumptions to aid in the temperature-extraction process, such as the surrounding temperature, the humidity and transmittance of the atmosphere, etc.

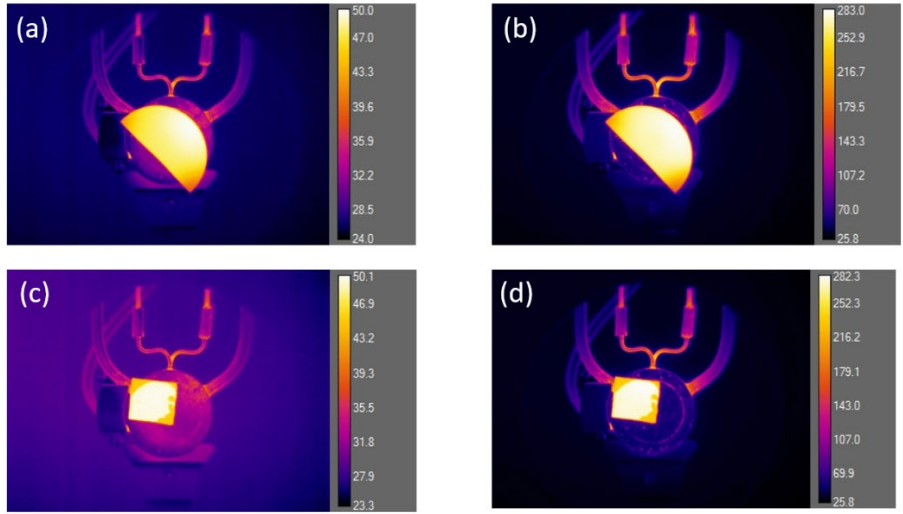

Figure. S4. Infrared camera image of the fused-silica window (a-b) and the CNT blackbody (c-d) when the heater temperature was set to 50 (left) and 300 °C (right). The color bar is the measured temperature, assuming $\epsilon_{set}$ of 0.89 for fused-silica window and 0.97 for the CNT blackbody.

## S4: Obtaining material properties from ellipsometry measurements

To extract the temperature from the measured thermal-emission spectrum, we needed precise values of $n(\lambda)$ and $\kappa(\lambda)$ of our sample at different temperatures. Therefore, we performed ellipsometry measurements on our fused-silica window with incident angles of 35, 45 and 55°, for free-space wavelengths from 4 to 15 $\mu$m, at 50, 200 and 300 °C. The complex refractive indices were then extracted by fitting the raw data ($\Psi$ and $\Delta$), assuming an infinitely thick sample with uniform temperature. The assumption of uniform temperature is reasonable considering the limited temperature gradient and the anticipated modest change in $n/\kappa$ as a function of temperature. Figure S5 shows the results, where the amplitude of the vibrational resonances near 9 $\mu$m becomes slightly smaller and the resonance width becomes slightly broader as the temperature increases. Away from the vibrational resonances, the changes in $n$ and $\kappa$ from room temperature to 300 °C are negligible.

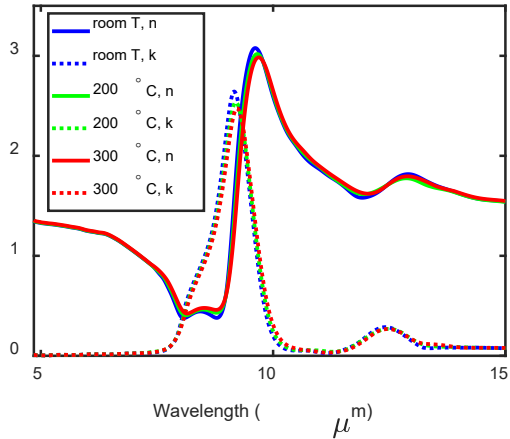

FIG. S5. Real (solid) and imaginary (dotted) parts of the refractive index of our 1-mm-thick fused-silica window, extracted using spectroscopic ellipsometry at room temperature (blue), 200 (green), and 300 ℃ (red).

## S5: Detailed description of the temperature-extraction process

### S5.1 Temperature extraction assuming no instrument limitations

In the ideal case without noise, the extraction of the temperature distribution from the thermal-emission spectrum is very robust because there is a unique combination of temperatures corresponding to a specific spectrum (*i.e.* there is a unique, exact solution of Eq. 4). We demonstrate this point numerically here.

As in the main text, we consider a 1-mm-thick fused-silica window, assuming its temperature drops linearly from 300 ℃ at the bottom surface to 283 ℃ at the top surface, and calculate its corresponding thermal emission spectrum [Fig. S6(a)]. In the calculation, the 1-mm-thick fused silica was modeled as an 11-layer structure, with each layer having the same optical material properties, but different temperatures.

Then, we used the calculated emission values at the 11 different wavelengths, and sent them into a nonlinear equation solver [lsqnonlin(fun, $x_0$) in Matlab] to solve the 11 different temperatures. This solver starts at $x_0$ and finds a minimum of the sum of squares of the functions in "fun", which is Eq. 4 in the main text. In solving temperature, we set an upper and lower bound of 330 and 250 ℃, respectively. At first, all 11 spectral points were chosen in the opaque region of fused silica [black dots, Fig. S6(a) for $\lambda > 8\ \mu m$]. In this case, the inversion process was only able to recover the temperature of the top layer [Fig. S6(b)], as expected. Then, all 11 spectral points were chosen in the semitransparent region [red dots, Fig. S6(a) for $\lambda > 8\ \mu m$], and all 11 temperatures were easily recovered from the spectral data [Fig. S6(c)]. In this extraction process, we did not need to make any assumptions about the shape of the temperature distribution.

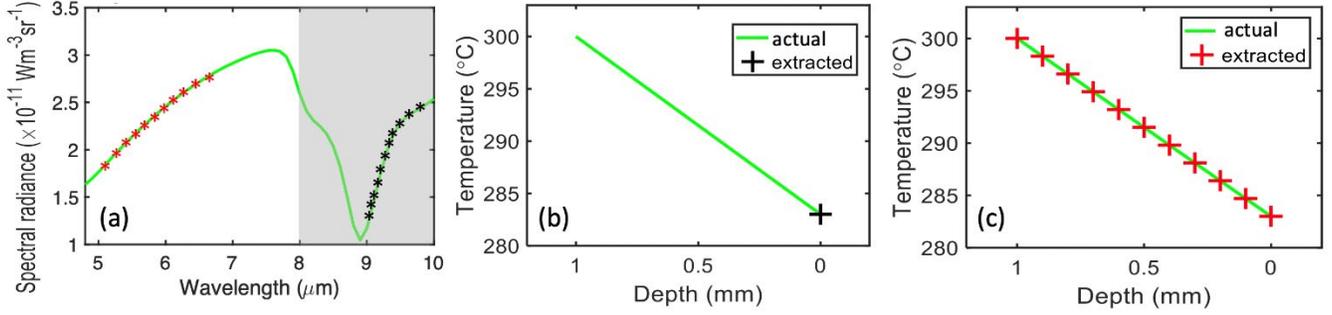

FIG. S6. (a): Calculated thermal-emission spectrum from a 1-mm-thick fused-silica window, assuming a linear temperature drop from 300 to 283 ℃ from the bottom surface to the top surface. (b) Recovered temperatures from the emission data taken from the opaque region [$\lambda > 8\ \mu m$, black dots in (a)] of fused silica. (c) The same as (b) but using the semitransparent region [$\lambda < 8\ \mu m$, red dots in (a)].

In the absence of noise, this inversion process works for arbitrary temperature profiles. As an example, the 1-mm-thick fused-silica window is assumed to have some arbitrary temperature distribution shown in Fig. S7(b). The corresponding thermal-emission spectrum is shown in Fig. S7(a). In this case, the nonlinear equation solver is still able to recover all the temperatures [Fig. S7(b)] from the calculated spectrum.

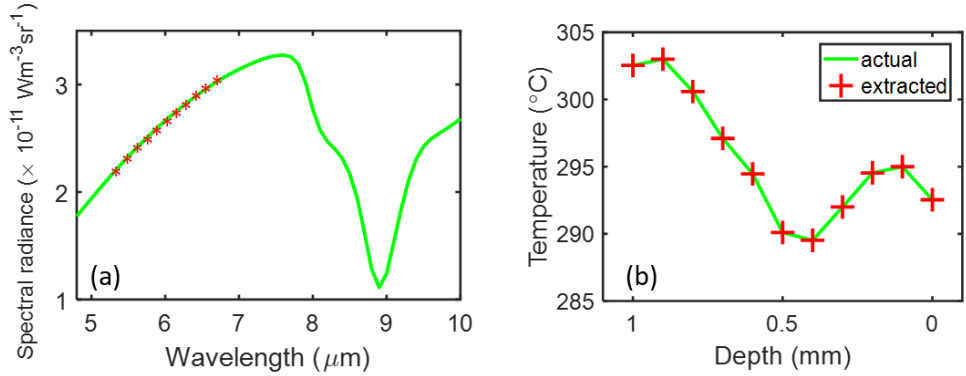

FIG. S7. (a): Calculated thermal-emission spectrum from a 1-mm-thick fused-silica window with an arbitrary temperature distribution along the vertical direction, as indicated by the green line in (b). Recovered temperature using the emission spectrum within the semitransparent spectral region of fused silica (red symbols).

*S5.2 Temperature extraction in realistic experimental conditions with noise*

In any experimental setting, the measured thermal-emission spectra are inevitably noisy. Extracting temperature distributions from noisy spectra is much more difficult. To demonstrate this, we calculated the emission spectrum with $\Delta\lambda = 100$ nm and added random fluctuations with relative amplitude of 1% to each wavelength points of the exact spectrum [*i.e.*, $I_{noisy}(\lambda_i) = I(\lambda_i)(0.99 + 0.2\text{rand})$, where "rand" returns a uniformly distributed random number in the interval (0,1)], as shown in Fig. S8 (a). The extracted temperatures using the nonlinear equation solver [i.e., the temperature profile that returns the minimum error in Eq. 4] in the bounded region between 240 and 350 °C using the noisy spectrum are plotted using red crosses in Fig. S8(b). As shown here, the extracted temperatures are drastically different from the input. The reason behind this is that the one-to-one relationship between temperature and thermal-emission spectrum breaks down when noise is present. In this case, there is no exact solution to Eq. 4 and there are many possible combinations of temperatures resulting emission spectra that have similar error with respect to the noisy spectrum. The extracted temperature distribution from the nonlinear equation solver, which tries to find the minimum error of all these possible combinations in a certain temperature range, depends very sensitively on the exact value of noise.

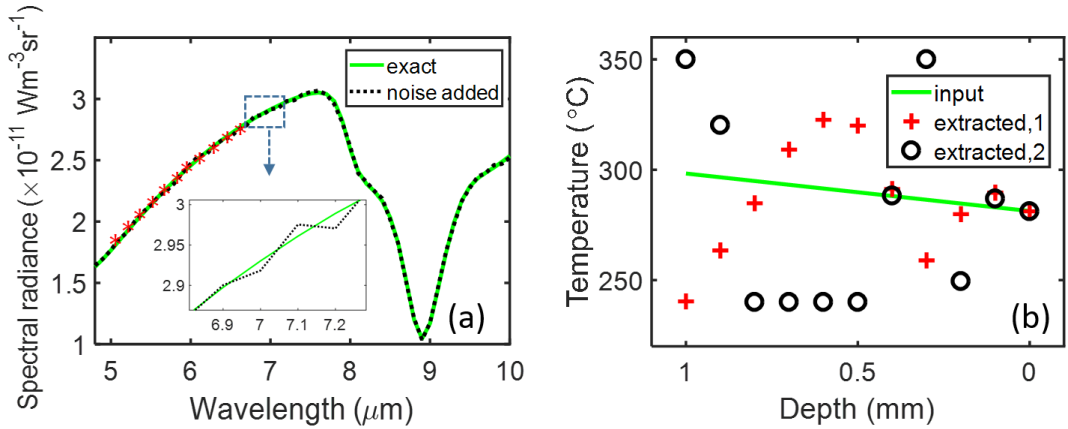

FIG. S8. (a): Calculated thermal-emission spectrum from a 1-mm-thick fused-silica window assuming a linear temperature drop from 300 to 283 °C from the bottom surface to the top surface, as indicated by the green line in (b). 1% of random noise is added into each wavelength point of the spectrum (black dotted line; also see inset). (b): Recovered temperature profiles from the noisy emission spectra within the semitransparent region of fused silica. The extractions from emission spectra with different random noise are quite different from one another, and from the input temperature distribution.

To further demonstrate this point, we generated the 1% random noise with a different random seed, and plotted the extracted temperatures using black circles in Fig. S8(b). The extracted temperatures look drastically different not only from the input linear distribution, but also from the temperatures extracted from the previous noisy spectrum with a different seed for the noise. Hence, extraction of temperature distribution from non-ideal thermal-emission spectra is not trivial.

One approach that enables a more robust temperature extraction is to use more spectral points ($M$) in the thermal-emission spectrum to solve the temperatures ($N$): i.e., $M > N$. One can understand why this works by considering the extreme case where there is an infinite number of wavelength points, but the noise level per point remains the same. In this case, the spectrum that minimizes the error in Eq. 4 would be the exact spectrum due to the random nature of the noise. If, on the other hand, $M$ is not big enough, the spectrum that has the minimum error can be different from the exact spectrum. Hence the extracted temperature can be different from the actual temperature. In reality, the noise level increases as the resolution increases because a finite signal is divided into more wavelength, leading to an increase of the relative noise level. Here we assume such relative noise level does not change with the number of "bins" for simplicity.

To demonstrate this point, we consider a relatively simple case: a two-layer model. In this simple model, the 1-mm-thick fused-silica window is assumed to be at 280 °C for the top half and 290 °C for the bottom half, with an abrupt transition at the interface. The local emissivity of the top and bottom layers were calculated using the scattering-matrix method. Then we added random noise to the spectrum and performed a brute-force (exhaustive) sweep of different combinations of $T_{top}$ and $T_{bottom}$ to find the combination that returns the global minimum error:

$$Error(T_{top}, T_{bottom}) = \sum_{\lambda_1}^{\lambda_2} \{I_{noisy}(\lambda) - \bar{\epsilon}_{top}(\lambda)I_{BB}(\lambda, T_{top}) - \bar{\epsilon}_{bottom}(\lambda)I_{BB}(\lambda, T_{bottom})\}^2 \quad \text{(S10)}$$

Error maps with four different random noise seeds and two different numbers of spectral points $M$ are shown in Fig. S9. In Fig. S9, $\lambda_1 = 4.8$ and $\lambda_2 = 5\ \mu m$ are fixed, and the wavelength points (*i.e.*, $\lambda$ in Eq. S10) are picked uniformly between $\lambda_1$ and $\lambda_2$ with $\Delta\lambda = 50$ nm for the top row and $\Delta\lambda = 0.1$ nm for the bottom row. Therefore, only 5 different wavelength points were used to generate the top panels, while 2000 wavelength points were used to generate the bottom panels. The magnitude of the noise added here is 1% for each of the 5 and 2000 wavelength points [*i.e.*, $I_{noisy}(\lambda_i) = I(\lambda_i)(0.99 + 0.2\text{rand})$].

As shown from the top panels in Fig. S9, the combination of $T_{top}$ and $T_{bottom}$ that returns the minimum error changes significantly when different random noise is added to the exact spectrum. This is because only 5 wavelength points are used. On the other hand, if enough wavelength points are used (bottom figures of Fig. S9), the combination of $T_{top}$ and $T_{bottom}$ that returns the minimum error overlaps with the input temperature and does not change when different random noise is added on top of the exact spectrum. Such temperature extraction is very robust.

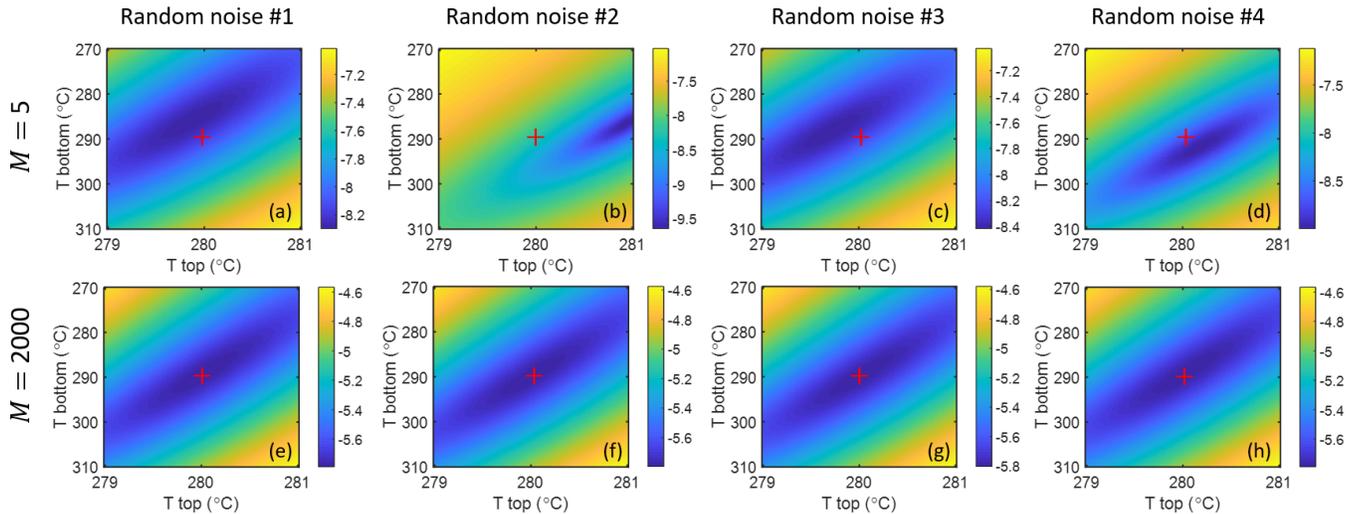

FIG. S9. (a-d): Distribution of $\log_{10}(Error)$ for different values of $T_{top}$ and $T_{bottom}$ from Eq. S9 (with $\lambda_1 = 4.8$, $\lambda_2 = 5\ \mu m$ and $\Delta\lambda = 50$ nm so that $M = 5$) for the same level of random noise per wavelength but with different random seeds. The temperatures that generate the exact spectrum are marked with red crosses. (e-h): same as (a-d) but with $\Delta\lambda = 0.1$ nm and $M = 2000$.

One can do similar analysis for more complicated cases, such as with three or more layers. As the layer number increases, the number of unknown temperatures that needs to be determined increases. Consequently, the number of wavelength point needed for robust temperature extraction for a fixed amount of noise per wavelength point increases as well. In other words, temperature extraction from a noisy spectrum must sacrifice resolution with robustness in the case of limited number of wavelength points. For the experimentally measured thermal-emission spectrum from the fused-silica window, we have about 450 wavelength points with a noise level of about 1% per wavelength point. We found that robust temperature extraction is possible for the experimental data assuming four layers [Fig. 4(c)]. When more layers are included (i.e., assuming a 5 or more-layer structure for the fused-silica window), temperature extraction becomes much more challenging.

One way to make temperature extraction much more robust without improving the experimental data is to put constraints onto the potential temperature distributions. For example, some functional form with a few parameters can be assumed to describe the temperature vs. depth. In this case, instead of solving for $N$ unknown independent temperatures, only the parameters of the assumed functional form need to be determined.

One example demonstrated in the main text is the case of a piece of fused silica on top of a heater. Assuming no radiative heat transfer, Fourier's law of heat conduction tells us that the temperature profile inside the fused silica is a linear function of depth when it is in a steady state:

$$T(z) = T_{top} + \alpha z \qquad (\text{S11})$$

Therefore, only two parameters need to be determined: $\alpha$ and $T_{top}$. As shown in [Fig. 4(c)], the extracted temperature from the experimental spectrum after making the assumption in Eq. S11 is very close to the true value.

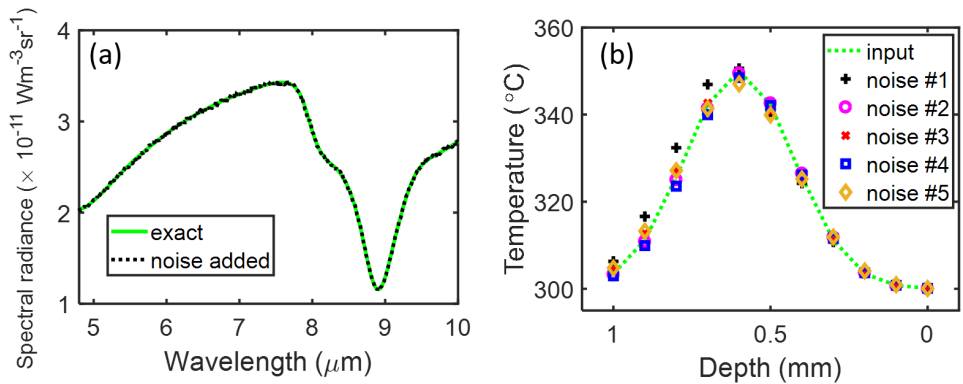

FIG. S10. (a): Calculated thermal-emission spectrum from a 1-mm-thick fused-silica window assuming a Gaussian temperature distribution along the depth direction, as indicated by the green line in (b). 1% of random noise is added into each wavelength points ($\Delta\lambda = 12.5$ nm) of the exact spectrum [i.e., $I_{noisy}(\lambda_i) = I(\lambda_i)(0.99 + 0.2\text{rand})$] (black dotted line). (b): Recovered

temperature profiles with different seeds of 1% random noise are shown by different markers.

As a further demonstration of this approach, we consider the case of Gaussian-like temperature distribution. Such a temperature distribution may be found in active devices such as the light-emitting diodes (LEDs) [S6] and quantum cascade lasers [S7], [S8]. Due to the electron-hole recombination process, the active layer of such devices is expected to be much hotter than the surrounding regions. As a numerical example, a 1-mm-thick fused silica window is assumed to have a Gaussian distribution in the depth direction:

$$T(z) = T_{top} + \alpha e^{-(z-z_0)^2/\beta} \tag{S12}$$

An input temperature distribution with the following parameters was assumed for the fused-silica window: $T_{top} = 300\ °C$, $\alpha = 50\ °C$, $z_0 = 0.4$ mm, and $\beta = 0.06$ mm$^2$. The temperature distribution is shown using the green line in Fig. S10(b) and the corresponding thermal-emission spectrum is shown using the green line in Fig. S10(a). In the calculation, we picked $\lambda_1 = 4.8$ and $\lambda_2 = 10\ \mu m$, with $\Delta\lambda = 12.5$ nm, so there were total of ~400 wavelength points. 1% of random noise was added onto each wavelength points of the calculated exact emission spectrum [$i.e., I_{noisy}(\lambda_i) = I(\lambda_i)(0.99 + 0.2\text{rand})$, black dotted line in Fig. S10(a)] and the noisy spectrum was then sent into the nonlinear equation solver to extract the temperature profile. In the extraction, even though the fused-silica window was divided into 11 layers, only 4 parameters needed to be determined. As shown in Fig. S10(b), the extracted temperature profile is very robust against noise, as the extraction does not change too much when the same level of noise with different random seeds was added into the spectrum.

# References

[S1]    G. Kirchhoff, "Über das Verhältnis zwischen dem Emissionsvermögen und dem Absorptionsvermögen der Körper für Wärme und Licht," in *Von Kirchhoff bis Planck*, Wiesbaden: Vieweg+Teubner Verlag, 1978, pp. 131–151.

[S2]    Y. Xiao *et al.*, "Measuring Thermal Emission Near Room Temperature Using Fourier-Transform Infrared Spectroscopy," *Phys. Rev. Appl.*, vol. 11, no. 1, p. 014026, Jan. 2019.

[S3]    R. Boyd, *Radiometry and the detection of optical radiation*. John Wiley & Sons, 1983.

[S4]    J. Kischkat *et al.*, "Mid-infrared optical properties of thin films of aluminum oxide, titanium dioxide, silicon dioxide, aluminum nitride, and silicon nitride," *Appl. Opt.*, vol. 51, no. 28, p. 6789, Oct. 2012.

[S5]    H. Ye, X. J. Wang, W. Lin, C. P. Wong, and Z. M. Zhang, "Infrared absorption coefficients of vertically aligned carbon nanotube films," *Appl. Phys. Lett.*, vol. 101, no. 14, p. 141909, Oct. 2012.

[S6]    M. Arik, C. A. Becker, S. E. Weaver, and J. Petroski, "Thermal management of LEDs: package to system,"

in *Third international conference on solid state lighting* 2004, vol. 5187, p. 64.

[S7]  V. Spagnolo *et al.*, "Temperature profile of GaInAs/AlInAs/InP quantum cascade-laser facets measured by microprobe photoluminescence," *Appl. Phys. Lett.*, vol. 78, no. 15, pp. 2095–2097, Apr. 2001.

[S8]  C. A. Evans, V. D. Jovanovic, D. Indjin, Z. Ikonic, and P. Harrison, "Investigation of Thermal Effects in Quantum-Cascade Lasers," *IEEE J. Quantum Electron.*, vol. 42, no. 9, pp. 857–865, Sep. 2006.